\documentclass[prb,reprint,showpacs,longbibliography]{revtex4-1}

\usepackage{graphicx}
\usepackage{amsfonts}
\usepackage{amsmath}

\usepackage[position=top]{subfig}

\DeclareGraphicsExtensions{.eps,.jpg}

\begin{document}

\title{Continuum limit of lattice models with Laughlin-like\\ ground states containing quasiholes}

\author{Iv\'an D. Rodr\'iguez}
\author{Anne E. B. Nielsen}
\affiliation{Max-Planck-Institut f\"ur Quantenoptik, Hans-Kopfermann-Stra{\ss}e 1, D-85748 Garching, Germany}

\begin{abstract}
There has been a significant interest in the last years in finding fractional quantum Hall physics in lattice models, but it is not always clear how these models connect to the corresponding models in continuum systems. Here we introduce a family of models that is able to interpolate between a recently proposed set of lattice models with Laughlin-like ground states constructed from conformal field theory and models with ground states that are practically the usual bosonic/fermionic Laughlin states in the continuum. Both the ground state and the Hamiltonian are known analytically, and we find that the Hamiltonian in the continuum limit does not coincide with the usual delta interaction Hamiltonian for the Laughlin states. We introduce quasiholes into the models and show analytically that their braiding properties are as expected if the quasiholes are screened. We demonstrate screening numerically for the 1/3 Laughlin model and find that the quasiholes are slightly smaller in the continuum than in the lattice. Finally, we compute the effective magnetic field felt by the quasiholes and show that it is close to uniform when approaching the continuum limit. The techniques presented here to interpolate between the lattice and the continuum can also be applied to other fractional quantum Hall states that are constructed from conformal field theory.
\end{abstract}


\maketitle

\section{Introduction}
Quantum many-body systems can display a wealth of peculiar phenomena that are interesting from both a fundamental and a practical point of view. An important example is topological systems with the possibility to have emergent particles with unusual properties like fractional charge and nontrivial braiding statistics. The fractional quantum Hall (FQH) effect plays a central role in this context since topological states were realized experimentally in these systems already in the early eighties\cite{FQH} and because the states appearing in the FQH effect are supposed to be well-described by relatively simple, analytical wave functions,\cite{laughlin,haldane,halperin,jain,M-R} which is a significant advantage for theoretical studies.

There is currently a lot of interest in investigating possibilities for obtaining and realizing the FQH effect in different systems. This research gives a deeper understanding of when and how the FQH effect can occur, and it can open up doors to investigate different aspects of the effect experimentally since different setups allow different properties to be measured. The hope is also to find particularly simple ways to realize the effect, which would significantly improve the possibilities for utilizing the effect in practical applications.

In the present paper, we are concerned with FQH models in lattice systems that are obtained by a strategy, in which one tries to keep the wave function as close as possible to the corresponding wave function in the continuum.\cite{schroeter,thomale,kapit,nielsen12,greiter,tu14,glasser} More specifically, one starts from an analytical FQH wave function in the continuum. A corresponding lattice FQH state is then defined by restricting the allowed positions of the particles to a set of lattice sites, and it may also be desirable to modify the state slightly. Finally, the analytical properties of the state are used to find a Hamiltonian for which the state is the unique ground state. It has turned out that conformal field theory (CFT) is a very helpful tool in this respect.\cite{M-R,nielsen11,nielsen12,tu14,glasser}

The Hamiltonians obtained by using this approach are typically valid only for a particular lattice filling factor. It is, however, interesting to ask if there is a way to directly connect the lattice models to the usual continuum models, since answering this question would give a more complete understanding of the lattice FQH models, would make it easier to compare lattice and continuum FQH models, and would provide further guidance on the possibilities for obtaining FQH models. In \onlinecite{tu14,glasser}, it has been shown how one can interpolate FQH wave functions between the lattice and the continuum limit within the context of CFT, but also in these papers a Hamiltonian was only found in the lattice limit. Interpolations from the continuum towards the lattice using an approximate Hamiltonian have been done as well,\cite{hafezi} but in this case the overlap between the ground state of the model and the (bosonic) Laughlin state is only high up to a moderate lattice filling factor, i.e.\ in the quasi-continuum limit.

Here, we propose an approach that allows us to obtain both lattice and continuum models with FQH ground states and to interpolate between the two limits within the same family of models. The models are built using CFT and allow us to compare the CFT Hamiltonian to the usual delta function interaction Hamiltonian in the continuum. We find that the two do not coincide. The construction also allows us to add localized quasiholes to the models and study their properties for different lattice filling factors. We find that the quasiholes are slightly better screened in the continuum than in the lattice and that they have the expected braiding properties.

The paper is structured as follows. In Sec.\ \ref{sec:model}, we introduce a family of wave functions and corresponding Hamiltonians. The wave functions are bosonic and fermionic Laughlin states on lattices for particular choices of the parameters. In Sec.\ \ref{sec:contlim}, we show that by choosing the parameters differently, it is possible within the same set of models to practically reproduce the continuous limit of the Laughlin states. This gives us a strategy to interpolate between the lattice and continuum Laughlin states in a way, where we know the Hamiltonian as well. In Sec.\ \ref{sec:interpolation}, we describe how we do the interpolation in practice, and we provide numerical results demonstrating the applicability of our method. The models also allow Laughlin quasiholes to be added, and we investigate their properties in Secs.\ \ref{sec:qh} and \ref{sec:braiding}. In particular, we show that the size of the quasiholes converges when going towards the continuum limit and that the quasiholes are slightly larger in the lattice than in the continuum. We also argue that the quasiholes have the expected braiding properties. In Sec.\ \ref{sec:magfield}, we compute the effective magnetic field seen by the quasiholes numerically. Finally, we conclude the paper with a discussion of the results in Sec.\ \ref{sec:conclusion}.

\section{Wave function and Hamiltonian}\label{sec:model}
In this section, we introduce the lattice model that we are considering. We first define a two-dimensional lattice with sites, in the complex plane, at the positions $z_j$, $j=1, \ldots ,N_0$. Laughlin-like states can be constructed\cite{tu14,nielsen15} on this lattice in terms of the vertex operators, $V_{n_j}(z_j)=(-1)^{(j-1)n_j}: \mbox{exp} [i(q n_j -1) \phi(z_j)/\sqrt{q}]:$, with $: \ldots :$ the standard normal ordering in Conformal Field Theory (CFT)\cite{CFTbook}, $\phi(z)$ a massless chiral free boson, $q$ a positive integer and $n_j \in \{ 0,1 \}$ the number of hard-core bosons/fermions at lattice site $z_j$ for $q$ even/odd. Quasihole excitations over the Laughlin states can be made at arbitrary complex positions $\eta_i$, $i=1, \ldots ,Q$, by means of the quasihole vertex operators, $W_{p_j}(\eta_j)={}: \mbox{exp} [i p_j \phi(\eta_j)/\sqrt{q}]:$, where $p_j$ is a positive integer that determines the charge of the quasihole as will be clear in a moment. The lattice wave functions are given by
\begin{eqnarray}
\left| \Psi_q \right> = \sum_{n_1 , \ldots , n_{N_0}} \Psi_q(\eta_{1\rightarrow Q}, n_{1\rightarrow N_0}) \left| n_1 , \ldots , n_{N_0} \right> ,
\label{gstate}
\end{eqnarray}
where $\eta_{1\rightarrow Q}$ is shorthand notation for $\eta_1 , \ldots ,\eta_Q$,
\begin{eqnarray}
\Psi_q(\eta_{1\rightarrow Q}, n_{1\rightarrow N_0}) && \propto \left< 0 \right| W_{p_1}(\eta_1) W_{p_2}(\eta_2) \ldots W_{p_Q}(\eta_Q) \nonumber \\
&& \times V_{n_1}(z_1) V_{n_2}(z_2) \ldots V_{n_{N_0}}(z_{N_0})  \left| 0 \right>
\label{CFT_gstate}
\end{eqnarray}
and $\left| 0 \right>$ is the CFT vacuum. Eq.~\eqref{CFT_gstate} evaluates to \cite{CFTbook}
\begin{multline}
\Psi_q(\eta_{1\rightarrow Q}, n_{1\rightarrow N_0}) = \\ \mathcal{C}(\eta_{1\rightarrow Q})^{-1} \delta_n
\prod_{i<j} (z_i - z_j)^{q n_i n_j}
\prod_{i<j} (\eta_i - \eta_j)^{p_i p_j/q}\\
\times\prod_{i,j} (\eta_i - z_j)^{p_i n_j} \prod_{i,j} (\eta_i - z_j)^{-{p_i/q}}
\prod_{i \neq j} (z_i - z_j)^{-n_i},
 \label{laugh}
\end{multline}
with $\mathcal{C}$ a real normalization constant and $\delta_n=1$ for
$\sum_{i=1}^{N_0} n_i =  (N_0 - \sum_{i=1}^Q p_i)/q$
and $\delta_n=0$ otherwise. The lattice filling fraction is defined as $\nu_{\textrm{Lat}}=(\sum_i n_i)/N_0$ and in the absence of quasiholes ($p_i=0$ for all $i$) it becomes the Landau level filling fraction, $\nu_{\textrm{Lat}}=\nu=1/q$. Note that from the definition of $\nu_{\textrm{Lat}}$ and using the previous constraint on $\sum_i n_i$ the charge of the $j$-th quasihole is given by $q_j = p_j/q$.

If the lattice is defined on a disc-shaped region and the area per lattice site, $a$, is the same for all sites, then it can be shown\cite{tu14} that the norm of the factors $\prod_{i,j} (\eta_i - z_j)^{-p_i/q}$ and $\prod_{i \neq j} (z_i - z_j)^{-n_i}$ approaches the usual Gaussian
factors $\mbox{exp} ( -\frac{1}{4}\frac{2 \pi}{a} \sum_{i=1}^{Q} \frac{p_i}{q} |\eta_i|^2)$ and $\mbox{exp} (- \frac{1}{4}\frac{2\pi}{a} \sum_{i=1}^{N_0} n_i |z_i|^2 )$ in the thermodynamic limit. The phase factors can be transformed away if desired, and the state \eqref{laugh} then looks very similar to the Laughlin state in the continuum limit, \cite{FQHE-book1,FQHE-book2}
\begin{eqnarray}
&&\Psi_q(\eta_{1\rightarrow  Q}, n_{1\rightarrow N_0}) = \mathcal{C'}(\eta_{1\rightarrow Q})^{-1} \delta_n
\prod_{i<j} (z_i - z_j)^{q n_i n_j}  \nonumber \\
&& \prod_{i,j} (\eta_i - z_j)^{p_i n_j}
e^{- \frac{1}{4}\frac{2\pi}{a} \sum_{i=1}^{N_0} n_i |z_i|^2 }
e^{-\frac{1}{4}\frac{2\pi}{a} \sum_{i=1}^{Q} \frac{p_i}{q} |\eta_i|^2} ,
\label{gauss_fac}
\end{eqnarray}
where we have absorbed in $\mathcal{C'}$ the products $\prod_{i<j} (\eta_i - \eta_j)^{p_i p_j / q}$ and $\prod_{i,j} (\eta_i - z_j)^{-p_i/q}$ that are independent of $n_i$. Indeed, the state \eqref{laugh} is a lattice version of the Laughlin state in the continuum, \cite{tu14,nielsen15} i.e.\ in the absence of quasiholes it has a uniform lattice density and its quasihole excitations have the same topological properties as the Laughlin quasiholes in the continuum. Note that from the exponentials in \eqref{gauss_fac} we can define the lattice magnetic length, $\ell=\sqrt{a/(2\pi)}$. \cite{Regnault_latt}

Using the results in \onlinecite{tu14,nielsen15}, it is possible to construct a Hamiltonian
\begin{eqnarray}
&& H_Q =\sum_{i=1}^{N_0} \Lambda_i^\dagger \Lambda_i + c \left[  \sum_{i=1}^{N_0} n_i - \frac{1}{q}(N_0 - \sum_{j=1}^Q p_j ) \right]^2 \qquad \mbox{with} \nonumber \\
&& \Lambda_i = \sum_{j(\ne i)}^{N_0} \frac{1}{z_i-z_j} \left[ d_j -d_i(q n_j -1)  \right] -
\sum_{j=1}^Q \frac{p_j}{z_i-\eta_j} d_i \nonumber \\
\label{latt_hamilt}
\end{eqnarray}
whose ground state is \eqref{laugh}. Here, $c$ is a positive constant, and we note that the second term in $H_Q$ simply fixes the number of particles in the ground state to $\sum_{i=1}^{N_0} n_i=(N_0-\sum_{j=1}^Q p_j)/q$. We find numerically for small systems that the ground state of $H_Q$ is unique for random choices of the lattice and quasihole coordinates.

\section{The continuum limit}\label{sec:contlim}
In order to approach the continuum limit, we need the number of lattice sites per area to increase, while keeping the number of particles per area constant. In \onlinecite{tu14,glasser}, it has been demonstrated that this can be achieved by replacing the vertex operators $V_{n_j}(z_j)$ by the vertex operators $V_{n_j}^{\xi}(z_j)=e^{i\pi(j-1)\xi n_j}: \mbox{exp} [i(q n_j -\xi) \phi(z_j)/\sqrt{q}]:$, where $\xi$ is a positive number. Increasing the number of lattice sites to $N=N_0+\Delta N$, this modifies the neutrality condition to $\sum_{i=1}^{N} n_i = (N\xi - \sum_{i=1}^Q p_{\eta_i})/q$. We can hence keep the number of particles constant by choosing $\xi=N_0/N$, and the continuum is obtained for $N\to\infty$.

Here, we show that we can obtain a continuum limit of the lattice model \eqref{laugh} and \eqref{latt_hamilt} by adding to it $\Delta N\gg N_0$ uniformly distributed extra lattice sites and the same number of uniformly distributed background quasiholes with charge $1/q$ without changing the area of the system. In this way, the number of particles $\sum_{i=1}^{N_0+\Delta_N} n_i = (N_0 - \sum_{i=1}^Q p_{\eta_i})/q$ again stays the same. The advantage of this approach compared to the other is that the Hamiltonian \eqref{latt_hamilt} remains valid. Note that the background quasiholes are placed at fixed positions with the only scope to reach the continuum limit, and they therefore play a different role than the \textit{physical} quasiholes. To easily distinguish the two, we shall use $\eta_j$ to denote the positions of the $Q$ physical quasiholes and $w_j$ to denote the positions of the $\Delta N$ background quasiholes.

We first discuss the continuum limit in the absence of physical quasiholes, i.e.\ $Q=0$. From Eqs.~\eqref{laugh} and \eqref{latt_hamilt}, in the presence of $\Delta N$ background quasiholes and $\Delta N$ extra sites, the lattice Hamiltonian and its ground state become
\begin{eqnarray}
&& H_{\Delta N} =\sum_{i=1}^N \Lambda_i^\dagger \Lambda_i + c \left[  \sum_{i=1}^N n_i - N_0/q  \right]^2 \qquad \mbox{with} \nonumber \\
&& \Lambda_i = \sum_{j(\ne i)}^N \frac{1}{z_i-z_j} \left[ d_j - d_i(q n_j -1)  \right] -
\sum_{j=1}^{\Delta N} \frac{1}{z_i-w_j} d_i \nonumber \\
\label{hamilt}
\end{eqnarray}
and
\begin{multline}
\Psi_{q, \Delta N}(w_{1\rightarrow \Delta_N}, n_{1\rightarrow N}) = \\ \mathcal{C}(w_{1\rightarrow \Delta_N})^{-1} \delta_n
\prod_{i<j} (z_i - z_j)^{q n_i n_j} \prod_{i<j} (w_i - w_j)^\frac{1}{q} \\  \times\prod_{i,j} (w_i - z_j)^{n_j} \prod_{i,j} (w_i - z_j)^{-\frac{1}{q}}
\prod_{i \neq j} (z_i - z_j)^{-n_i}.
\label{qhole}
\end{multline}
Note that the factors $\prod_{i,j} (w_i - z_j)^{n_j}$ and $\prod_{i \neq j} (z_i - z_j)^{-n_i}$ can be written as
\begin{eqnarray}
&&\prod_{i,j} (w_i - z_j)^{n_j} \prod_{i \neq j} (z_i - z_j)^{- n_i} = \nonumber \\
&& e^{ \sum_j \sum_i n_j \ln |w_i - z_j|}
e^{-\sum_i \sum_{j \ne i} n_i \ln |z_i - z_j|} e^{i \alpha},
\label{gaussian0}
\end{eqnarray}
with $\alpha$ a phase factor that can be transformed away from the wave function if desired. In the following, we assume that both the lattice sites and the background quasiholes are distributed uniformly. We shall use the term {\it unit cell} to refer to the unit cell of the original lattice with $\Delta N=0$, and we shall take $a$ to mean the area of this unit cell. In the limit $\Delta N \rightarrow \infty$, the sums in the last member of \eqref{gaussian0} can be written as
$\sum_j \sum_i n_j \ln |w_i - z_j| \rightarrow \rho_w \sum_j n_j \int d w \ln |w - z_j|$ and
$\sum_i \sum_{j \ne i} n_i \ln |z_i - z_j| \rightarrow \rho_z \sum_i \int d z n_i \ln |z_i - z|$ with $\rho_z=n_z/a$ and $\rho_w=n_\omega/a$ being the densities of sites and quasiholes and $n_z$ and $n_w$ the corresponding number of sites and quasiholes within one unit cell. Assuming in addition that the lattice is defined in a disc-shaped region, Eq.~\eqref{gaussian0} turns into,
\begin{eqnarray}
&&\prod_{i,j} (w_i - z_j)^{n_j} \prod_{i \neq j} (z_i - z_j)^{- n_i} = e^{\rho_w \sum_{j=1}^N \int d w n_j \ln |w - z_j|} \nonumber \\
&& e^{-\rho_z \sum_{i=1}^N \int d z n_i \ln |z_i - z|} e^{i \alpha} = e^{ 2\pi(\rho_w -\rho_z) \sum_{i=1}^N n_i \frac{|z_i|}{4}^2} e^{i \alpha}.
\nonumber \\
\label{gaussian}
\end{eqnarray}
>From \eqref{gaussian} it is clear that in order to obtain the right Gaussian factors (see Eq.~\eqref{gauss_fac}) in the continuum limit we must require that $\rho_z=\rho_w+1/a$, i.e.\
\begin{eqnarray}
n_z=n_w+1.
\label{den_constr}
\end{eqnarray}
This equation states that the number of background quasiholes must be equal to the number of extra lattice sites. Finally, because the factors $\prod_{i<j} (w_i - w_j)^{1/q}$ and $\prod_{i,j} (w_i - z_j)^{-1/q}$ are independent of $n_i$ they can be absorbed into the normalization constant $\mathcal{C}$ and we find that the state \eqref{qhole} can be written as the state \eqref{gauss_fac}, with $p_i=0$ for all $i$, but with the lattice sites covering the full space. Therefore, it is a continuum version of the lattice Laughlin state \eqref{laugh}-\eqref{gauss_fac}. From here on we will use the terms lattice and continuum limit for $\Delta N=0$ and $\Delta N/N_0\gg 1$ respectively.

The Laughlin states in the continuum are known to be the exact and unique ground state of the Hamiltonian
\begin{eqnarray}
H=H_0 + V^{(q)},
\label{Hamilt_Kivel}
\end{eqnarray}
with $H_0$ the Landau level Hamiltonian and with $V^{(q)}$ the Trugman-Kivelson short-range potential interaction given by,\cite{Trug-Kivel}
\begin{eqnarray}
V^{(q)}=\sum_{i<j}^N \left( \nabla_i^2  \right)^{(q-1)/2} \delta^{(2)}(z_i-z_j)
\label{pot_1}
\end{eqnarray}
for $q$ odd and
\begin{eqnarray}
V^{(q)}=\sum_{i<j}^N \left( \nabla_i^2  \right)^{(q-2)/2} \delta^{(2)}(z_i-z_j)
\label{pot2}
\end{eqnarray}
for $q$ even. However it is important to remark that our Hamiltonian \eqref{hamilt} is non-local and therefore very different in nature from the local one in Eq.\eqref{Hamilt_Kivel}.

Finally, we can construct in a similar way a set of physical quasiholes $\eta_1, \ldots , \eta_Q$  in the continuum limit described by the ground state
\begin{eqnarray}
\Psi_{q, \Delta N, Q} \sim \Psi_{q, \Delta N} \prod_{i,j} (w_i - \eta_j)^\frac{1}{q} \prod_{i<j} (\eta_i - \eta_j)^\frac{p_i p_j}{q} \nonumber \\  \times\prod_{i,j} (\eta_i - z_j)^{p_i n_j} \prod_{i,j} (\eta_i - z_j)^{-\frac{p_i}{q}}
\label{qh_gstate}
\end{eqnarray}
of the Hamiltonian
\begin{eqnarray}
&&H_{\Delta N, Q} =\sum_{i=1}^N \tilde{\Lambda}_i^\dagger \tilde{\Lambda}_i + c \left[  \sum_{i=1}^N n_i - \frac{N_0-\sum_{j=1}^Q p_j}{q}  \right]^2  \quad
\mbox{with} \nonumber \\
&&\tilde{\Lambda}_i = \Lambda_i - \sum_{j=1}^Q \frac{1}{z_i-\eta_j} d_i
\label{qh_hamilt}
\end{eqnarray}
and $\Lambda_i$ defined in \eqref{hamilt}.

\section{Construction to obtain a roughly uniform particle density} \label{sec:interpolation}
As shown in the previous section, to reach the continuum limit we have to add to the original lattice an additional set of $\Delta N\gg N_0$
uniformly distributed lattice sites and the same number of uniformly distributed quasiholes so that \eqref{den_constr} is fulfilled.
There are of course many different ways to reach such a limit. Naively one could consider, for instance, a random distribution of lattice sites and quasiholes satisfying \eqref{den_constr}. However, because from a computational point of view it is not possible to have a perfectly uniform distribution, many of the lattice sites will be close to the quasihole positions and the probability to find a particle on those lattice sites will be very small. This means that the lattice density per unit cell will be nonuniform and indeed we have checked that it is nonvanishing only in a small region of the cells.
\begin{figure}[htb]
\includegraphics[width=\columnwidth]{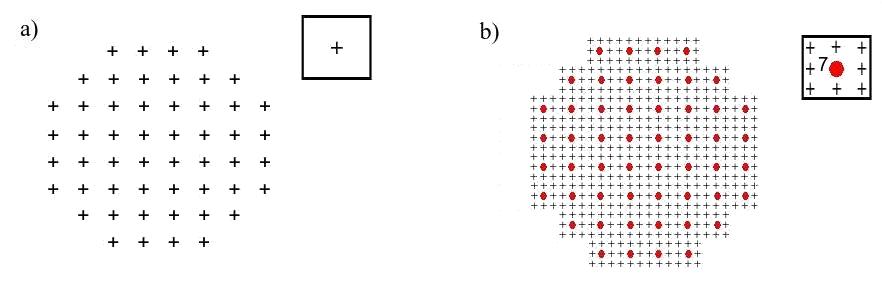}
\caption{{\small a) Lattice limit ($\Delta N_c=0$) and unit cell. b) Poor continuum limit approximation ($\Delta N_c=7$): unit cell made of 8 sites and 7 quasiholes (as indicated inside the unit cell). }}
\label{fig1}
\end{figure}
\begin{figure}[htb]
\includegraphics[width=\columnwidth]{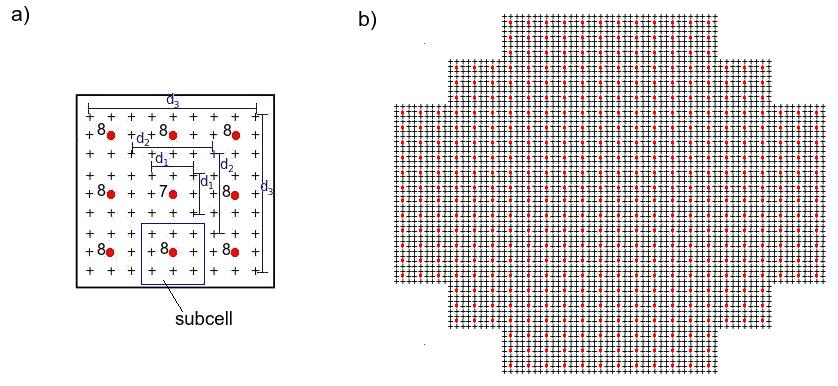}
\caption{{\small Continuum limit approximation ($\Delta N_c=71$) : a) Unit cell made of 9 subcells with the same structure as the unit cell in Fig.~\ref{fig1}b. All the subcells contain 8 quasiholes except for the one in the center with 7 quasiholes (note that this setting satisfies the constraint $n_z=n_w+1$). The $d_1$, $d_2$ and $d_3$ parameters are tuned by a Metropolis algorithm in order to reach an approximately uniform density in every subcell. b) Full lattice made of the unit cell in a).}}
\label{fig2}
\end{figure}
To overcome this problem and in order to interpolate between the lattice and the continuum limit we propose here to consider the lattice structure in Fig.~\ref{fig1}b with 8 sites distributed along the sides of the unit cell and 7 quasiholes all of them placed at the center  of the unit cell (note that $n_z=n_w+1$). This is clearly a poor approximation of the continuum limit because the number of extra lattice sites in a single unti cell, $\Delta N_c$, is very small ($\Delta N_c=7$). However, this construction can now be used as a building block to reach the continuum limit, as shown in Fig.~\ref{fig2}, where now the unit cell is made of $3 \times 3$ subcells, each of them with the same structure of the unit cell of Fig.~\ref{fig1}b. Note that in order to satisfy the constraint \eqref{den_constr} we keep the same number of quasiholes and lattice sites in all the subcells, except in the central one, that contains 7 quasiholes and 8 sites as shown in Fig.~\ref{fig2}a.
We also introduce, in every unit cell, a set of parameters $d_1$, $d_2$ and $d_3$ (see Fig.~\ref{fig2}a) to be tuned by a Metropolis algorithm, in order to better screen the effect of having one quasihole less in the subcell in the center than in the other subcells and thereby reach a more uniform density.
\begin{figure}[htb]
\includegraphics[width= 4.2cm, height=3.7cm]{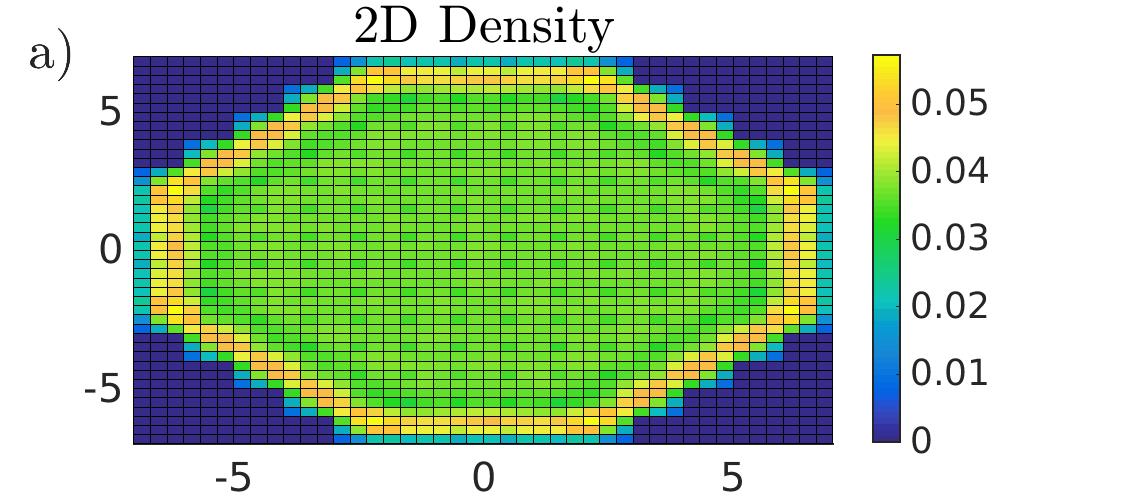}
\includegraphics[width= 4.35cm, height=3.7cm]{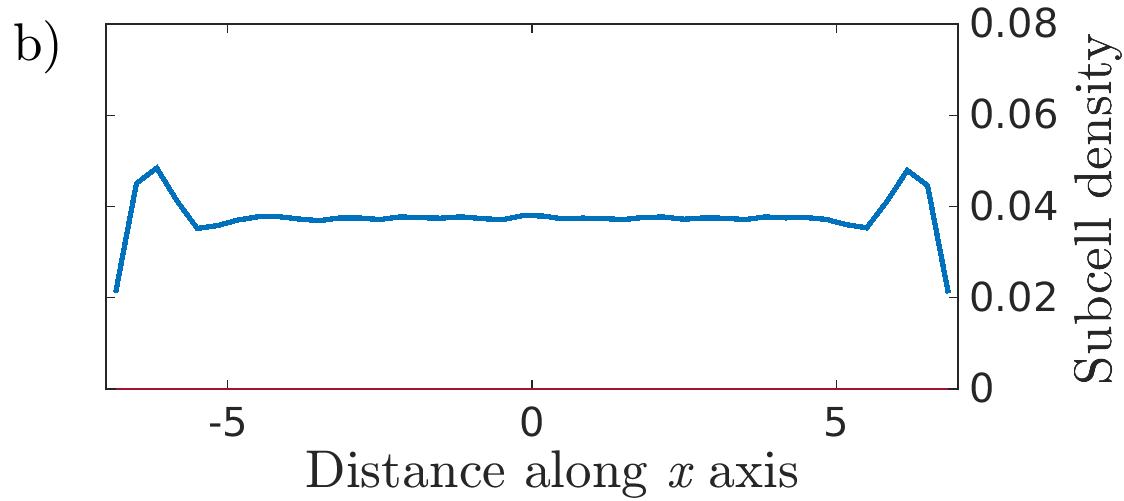}
\includegraphics[width= 4.2cm, height=3.5cm]{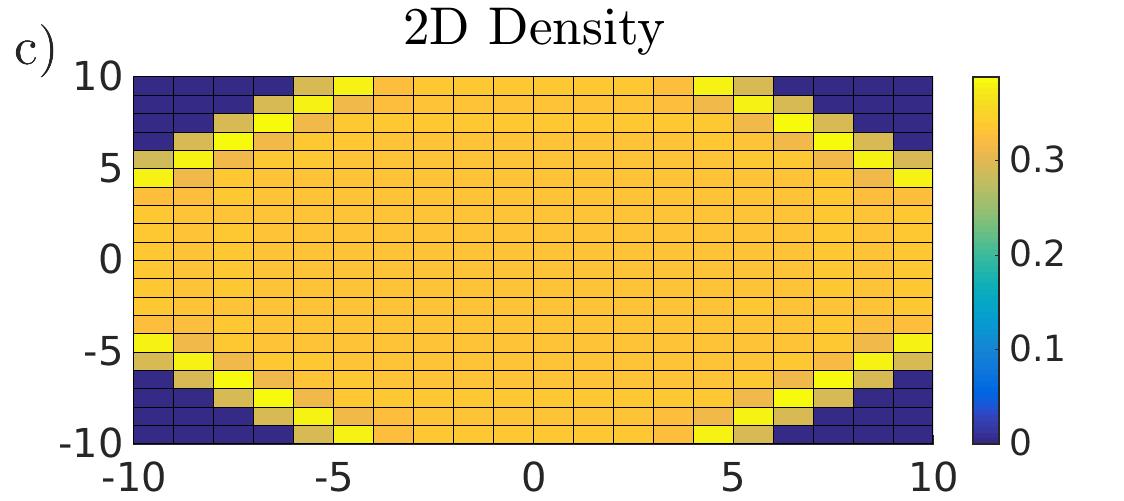}
\includegraphics[width= 4.35cm, height=3.7cm]{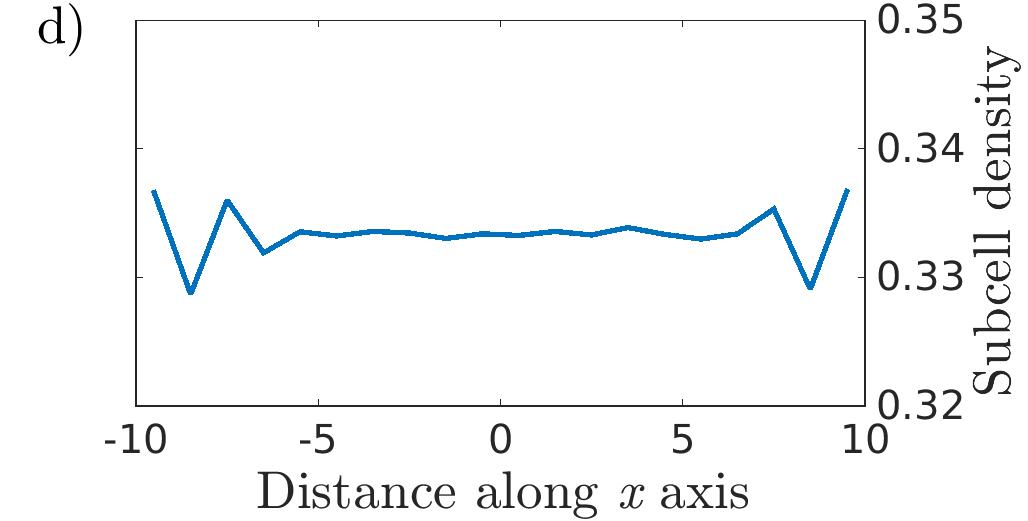}
\caption{{\small a) Two-dimensional density as defined in eq.\eqref{subcell_dens} for the $3 \times 3$ subcell case of Fig.~\ref{fig2} ($\Delta N_c=71$) and for $\nu=1/3$.
                 b) Subcell density along the x-axis of the plot in a). The density is almost uniform with small oscillations around the mean value
                    $\nu/9 \sim 0.0370 $.
                  The tuning parameters are given (in a=1 unities) by: $d_1=2/9+0.041087$, $d_2=4/9-0.011062$ and $d_3=8/9-0.001649$. c-d) Same plots as in a)-b) but for the lattice limit.}}
\label{fig3}
\end{figure}
In Fig.~\ref{fig3}a we show, for the $3 \times 3$ subcell case and $\nu=1/3$, the two-dimensional subcell density defined as,
\begin{eqnarray}
n(x_\alpha,y_\beta)=\sum_{i \in (x_\alpha,y_\beta)} \langle n_i \rangle,
\label{subcell_dens}
\end{eqnarray}
with $(x_\alpha,y_\beta)$ a set of coordinates pointing to the subcells and where the sum runs over all lattice sites $i$ within the $(x_\alpha,y_\beta)$ subcell (note that in the lattice limit eq.\eqref{subcell_dens} becomes the usual lattice density). Observe that the density is very close to uniform. This is more clear from Fig.~\ref{fig3}b where the subcell density
along the x-axis presents small oscillations around its mean value $\nu/9$ (the subcell density is given by $\nu$ divided by the number of subcells per unit cell). In Fig.~\ref{fig3}c-d we plot the same densities but for the lattice limit obtaining the expected results.\cite{nielsen15}
As we will show later a good approximation of the continuum limit is given by a unit cell made of $5 \times 5$ subcells, i.e. a unit cell with $\Delta N_c=199$ sites and 5 tuning parameters, chosen in a similar way as in Fig.~\ref{fig2}a. The density for this case and for $\nu=1/3$ is shown in Fig.~\ref{fig4} with the mean density given by $\nu/25$.
Note that the density oscillations (Fig.~\ref{fig4}b) are higher than for the $3 \times 3$ subcell case (Fig.~\ref{fig3}b) because we have to screen a much larger number of quasiholes with a few tuning parameters. This could be improved by including more parameters but, as it is clear from Fig.~\ref{fig4}, with this minimal setting the oscillations are relatively small and the density of the system is close enough to the uniform case.
\begin{figure}[htb]
\includegraphics[width= 4.25cm, height=3.7cm]{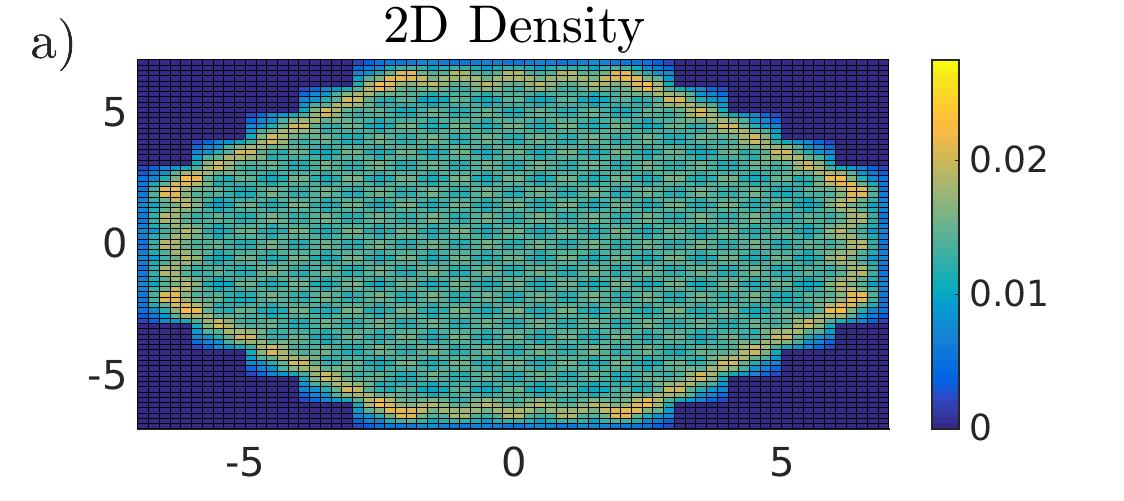}
\includegraphics[width= 4.25cm, height=3.7cm]{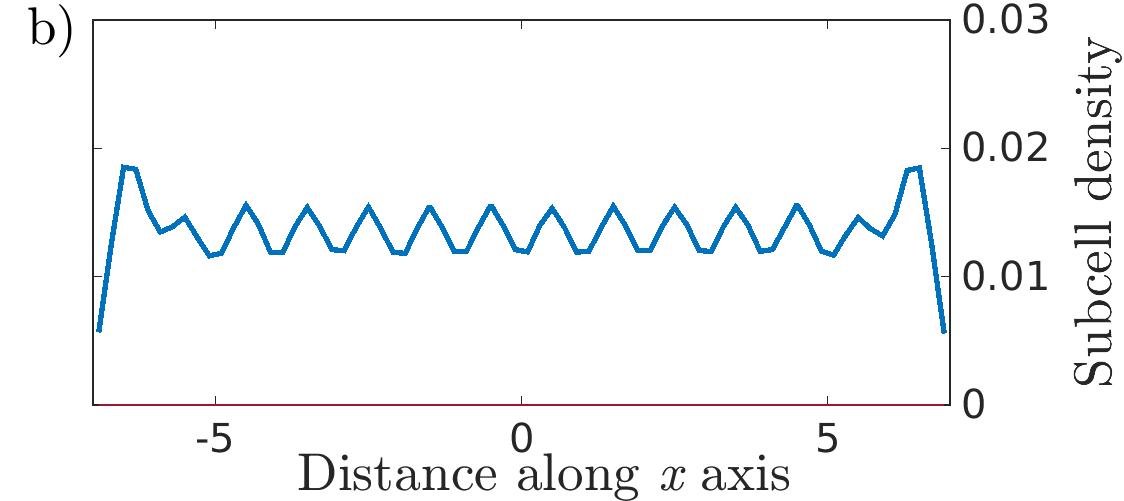}
\caption{{\small a) Two-dimensional density as defined in eq.\eqref{subcell_dens} for the $5 \times 5$ subcell case ($\Delta N_c=199$) and for $\nu=1/3$.
                 b) Subcell density along the x-axis of the plot in a). The density is close to uniform with small oscillations around the mean value  $\nu/25 \sim  0.0133 $. The tuning parameters are given (in a=1 unities) by: $d_1 =2/15-0.012522$, $d_2= 4/15 - 0.011956$, $d_3=8/15-0.002250$, $d_4=10/15-0.000954$ and $d_5=14/15+0.000027$.}}
\label{fig4}
\end{figure}

\section{Quasihole radius and charge}\label{sec:qh}
To study the extent of the quasiholes and their charge we split one particle into two quasiholes of charges $1/q$ and $(q-1)/q$ and we place them far away from each other.
To verify that the quasihole charge is equal to $1/q$ we compute the excess charge \cite{Regnault_latt, Papic_latt}
\begin{eqnarray}
&& {\bf Q}(r_j)= \sum_{i=0}^{j} \left[ \rho_\textrm{qh} (r_i) - \rho(r_i) \right] \mbox{; } r_i=(i+\frac{1}{2}) \epsilon, \mbox{ } i=0,1,2, \ldots
\nonumber \\
\label{ex_charge}
\end{eqnarray}
with $\rho_\textrm{qh}(r_i)$ the radial density at distance $r_i$ of one of the quasiholes,
\begin{eqnarray}
 && \rho_\textrm{qh} (r_i)  = \sum_{l \setminus r_i-\frac{\epsilon}{2} \le |z_l - \eta| <r_i + \frac{\epsilon}{2}} \left< n_l
 \right>_\textrm{qh} ,
\label{rad_dens}
\end{eqnarray}
$\left< n_l \right>_\textrm{qh}$ the quasihole mean density at position $z_l$, $\epsilon$ a small real number, $\eta$ the quasihole position and with $\rho(r_i)$ in Eq.~\eqref{ex_charge} given by \eqref{rad_dens} but replacing $\left< n_l \right>_\textrm{qh} \rightarrow \left< n_l \right>$.
\begin{figure}[htb]
\includegraphics[width= 4.5cm, height=3.5cm]{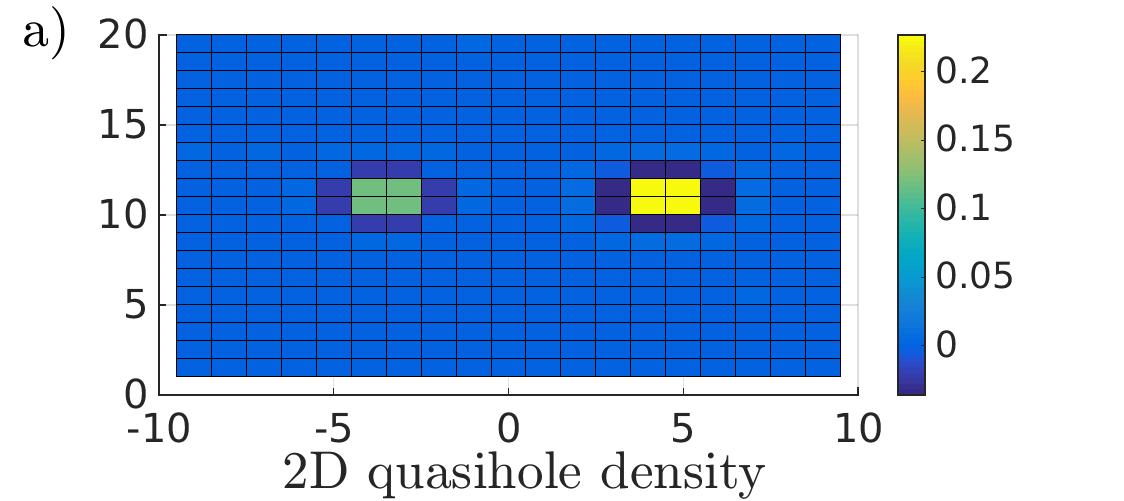}
\includegraphics[width= 4cm, height=3.5cm]{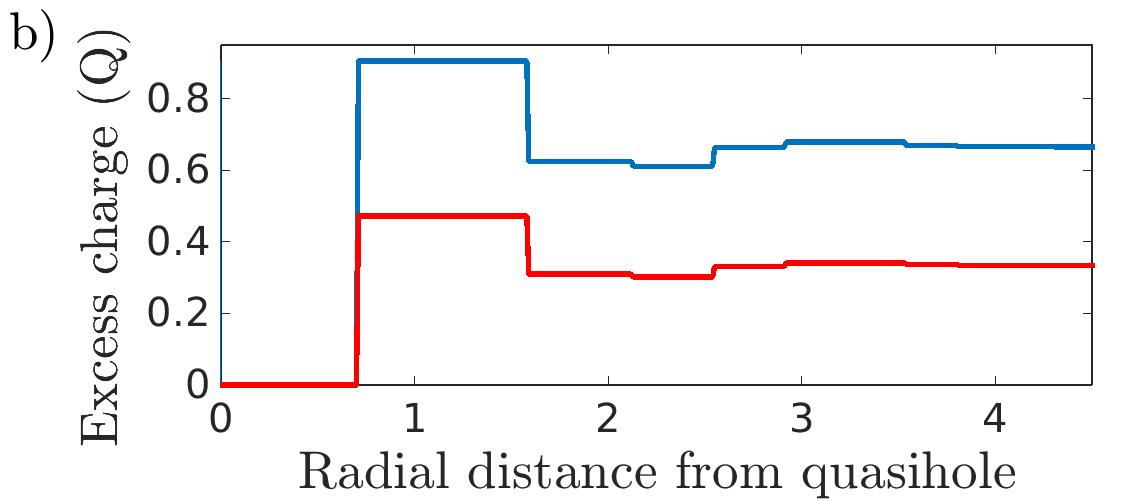}
\includegraphics[width= 4.5cm, height=3.5cm]{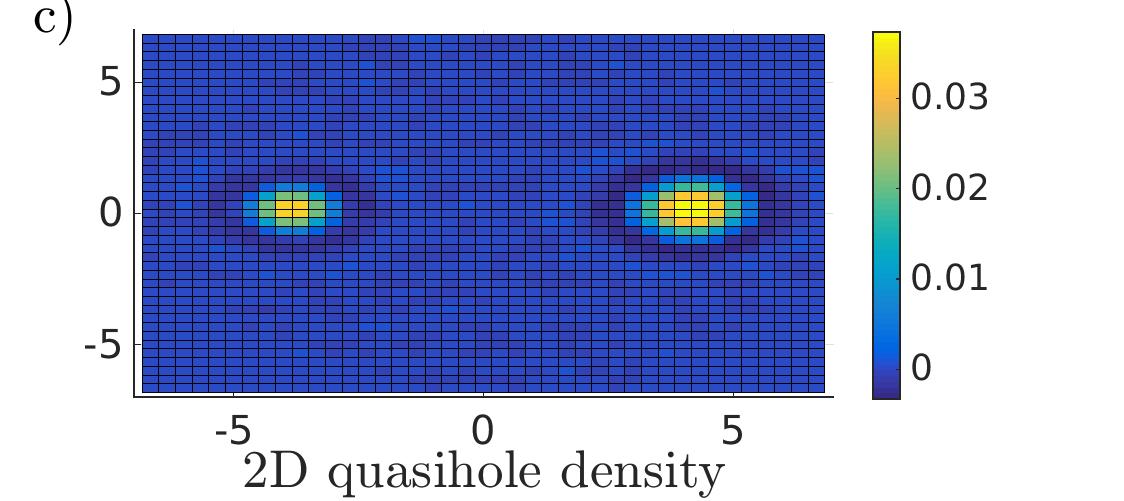}
\includegraphics[width= 4cm, height=3.5cm]{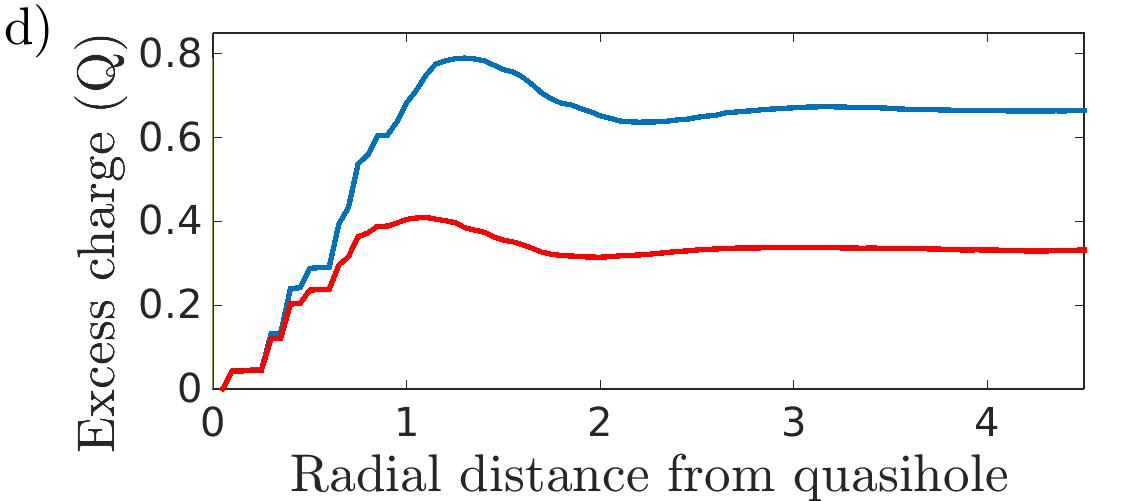}
\includegraphics[width= 4.5cm, height=3.5cm]{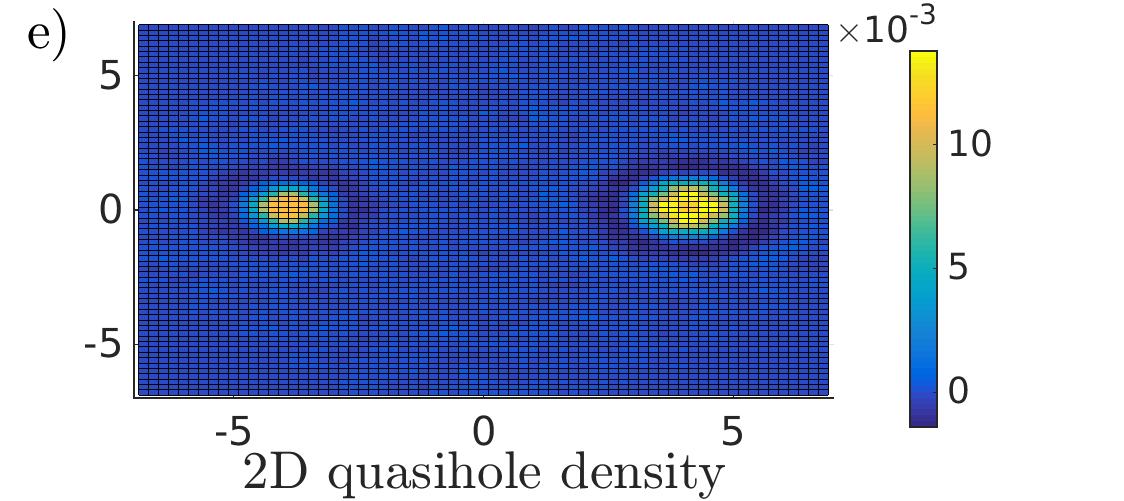}
\includegraphics[width= 4cm, height=3.5cm]{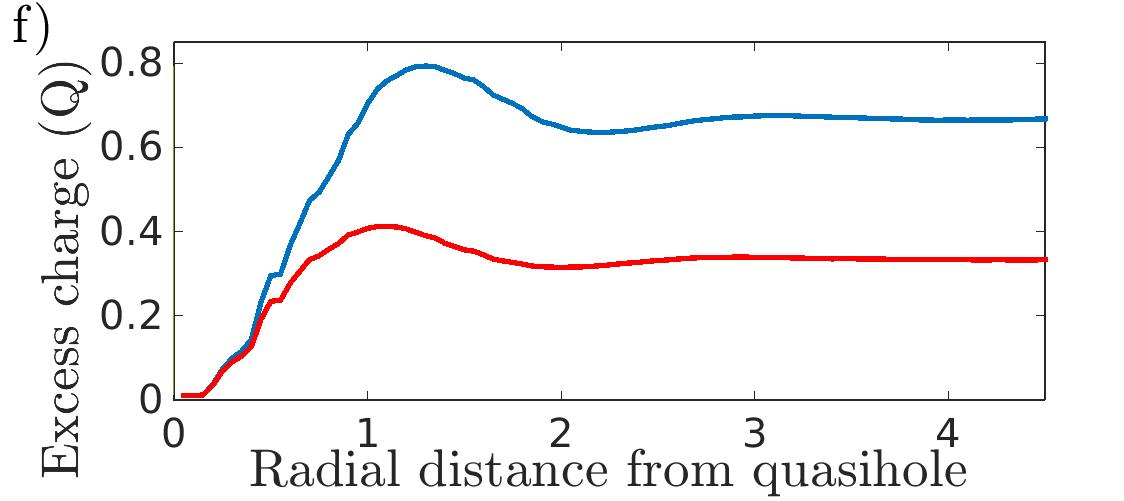}
\caption{{\small
a,c,e): Quasihole subcell density, $n_\textrm{qh}(x_\alpha,x_\beta) - n(x_\alpha,x_\beta)$, for the lattice limit (a) and the  $3 \times 3$ (c) and $5 \times 5$ (e) subcell cases and for $\nu=1/3$ quasiholes with charges $1/3$ and $2/3$ placed at positions $\eta_1=-4$ and $\eta_2=4$, respectively.
b,d,f): Excess charge, $\bf{Q}$, of both quasiholes, for $\nu=1/3$ and again for the lattice limit (b) and the $3 \times 3$ (d) and $5 \times 5$ (f) subcell cases. Note that $\bf{Q}$ tends to the quasihole charges $1/3$ (red curve) and $2/3$ (blue curve), as expected.}}
\label{fig5}
\end{figure}

In Fig.~\ref{fig5}a,c,e we show, for $\nu=1/3$ and for the lattice limit and the $3 \times 3$ and $5 \times 5$ cases, the subcell density,
$n_\textrm{qh}(x_\alpha,x_\beta) - n(x_\alpha,x_\beta)$ (see Eq.~\eqref{subcell_dens}), with $n_\textrm{qh}(x_\alpha,x_\beta)$ the subcell density including two quasiholes with charges $1/3$ and $2/3$ at positions $\eta_1=-4$ and $\eta_2=4$, respectively. From the figures it is clear that there is a very good screening of quasiholes for different values of $\Delta N_c$. In Fig.~\ref{fig5}b,d,f we place the quasiholes far away from each other and we show, again for $\nu=1/3$ and for the previous cases, that the excess charge, $\bf{Q}$, tends to the expected quasihole charges $1/3$ and $2/3$, respectively.

To compute the quasihole radius, $R_\textrm{qh}$, we use the second moment of $\rho(r)$, \cite{Regnault_latt, Papic_latt}
\begin{eqnarray}
R_\textrm{qh} = \sqrt{\frac{\sum_{i=0}^{j_\textrm{max}} \left| \rho_\textrm{qh} (r_i) - \rho(r_i) \right| r_i^2}{\sum_{i=0}^{j_\textrm{max}} \left| \rho_\textrm{qh} (r_i) - \rho(r_i) \right| }},
\label{radii}
\end{eqnarray}
with $j_\textrm{max}$ defined in such a way that $r_{j_\textrm{max}}$ is a distance far away from any quasihole, where $\rho_\textrm{qh}(r_{j_\textrm{max}}) -\rho(r_{j_\textrm{max}})$ vanishes.
In table \ref{table1}-\ref{table2} we show the values of $R_\textrm{qh}$ in magnetic length units, $\ell$, for the quasiholes introduced previously with charges $1/3$ and $2/3$ at $\nu=1/3$ and for the lattice limit and the $3\times3$ and $5\times5$ subcell cases.
Observe that $R_\textrm{qh}$ decreases with $\Delta N_c$ because of the better screening when one approaches the continuum limit (see Fig.~\ref{fig5}) and that the values of $R_\textrm{qh}$ for the $3\times3$ and $5\times5$ cases are very close, indicating that the $5 \times 5$ subcell case is a good approximation for the continuum limit.

Finally we want to remark that the values we obtain for $R_\textrm{qh}$ in the continuum limit are very close to the ones obtained in
the standard FQH effect \cite{bernevig,Regnault_latt} and other systems containing Laughlin-like quasihole states like Fractional Chern Insulators.\cite{Papic_latt}

\begin{table}
\begin{ruledtabular}
\begin{tabular}{@{}lll@{}}
Lattice ($\Delta N_c=0$) & $3 \times 3$ ($\Delta N_c=71$) & $5 \times 5$ ($\Delta N_c=199$)\\  \hline
3.0(3) & 2.0(3) & 2.0(5)
\end{tabular}
\end{ruledtabular}
\caption{\small{Quasihole radius, $R_\textrm{qh}$, for quasiholes with charges $1/3$ at $\nu=1/3$ and for different values of $\Delta N_c$.}}
\label{table1}
\end{table}

\begin{table}[htb]
\begin{ruledtabular}
\begin{tabular}{@{}lll@{}}
Lattice ($\Delta N_c=0$) & $3 \times 3$ ($\Delta N_c=71$) & $5 \times 5$ ($\Delta N_c=199$)\\  \hline
3.0(1)  & 2.3(5) & 2.3(5) \\
\end{tabular}
\end{ruledtabular}
\caption{\small{Quasihole radius, $R_\textrm{qh}$, for quasiholes with charges $2/3$ at $\nu=1/3$ and for different values of
$\Delta N_c$.
}}
\label{table2}
\end{table}

\section{Quasihole braiding}\label{sec:braiding}
In this section we show that the anyonic statistics \cite{FQHE-book1,FQHE-book2} remains invariant in the interpolation between the lattice and the continuum limit.

\begin{figure}[htb]
\includegraphics[width=0.9\columnwidth, viewport=200 378 435 450,clip]{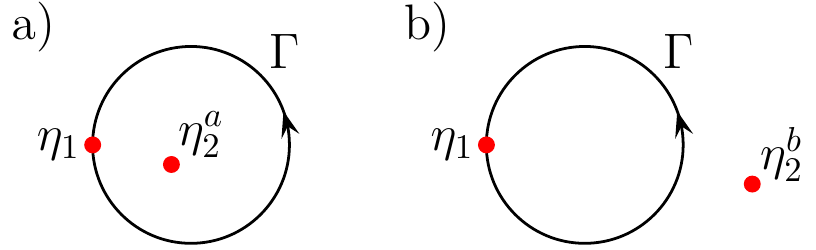}
\caption{{\small{Set up for anyonic statistic computation of quasiholes $\eta_1$ and $\eta_2$ with charges $\frac{1}{q}$ and $\frac{q-1}{q}$. In a) the quasihole $\eta_2$ is placed inside the curve $\Gamma$ (at position $\eta_2^a$) and in b) far away from it (at position $\eta_2^b$).}}}
\label{fig8}
\end{figure}

To compute the anyonic statistics we proceed as in the previous section and we split up one of the particles into
two quasiholes $\eta_1$ and $\eta_2$, with charges $1/q$ ($p_{\eta_1}=1$) and $(q-1)/q$ ($p_{\eta_2}=q-1$).
The anyonic statistics, $\gamma$, is defined in terms of the Berry phase \cite{Berry1,Berry2} and monodromy \cite{M-R} (i.e. the change obtained from
analytical continuation of \eqref{qh_gstate} when the quasiholes move around) as,
\begin{eqnarray}
2\pi \gamma = \theta_a + arg(M_a) - \theta_b - arg(M_b),
\label{alpha}
\end{eqnarray}
with $\theta_a$ and $M_a$ the Berry phase and monodromy due to the braiding of the quasiholes (Fig.~\ref{fig8}a) and $\theta_b$ and $M_b$ the same quantities but without quasihole braiding (Fig.~\ref{fig8}b).
The Berry phase for the state \eqref{qh_gstate} can be computed directly from the normalization constant
$\mathcal{C}(\eta_1,\eta_2, w_{1\rightarrow \Delta_N})$, \cite{nielsen15}
\begin{eqnarray}
\theta=\frac{i}{2} \int_{\Gamma} \frac{1}{\mathcal{C}^2} \frac{\partial{\mathcal{C}^2}}{\partial {\eta_1}} d \eta_1 -
\frac{i}{2} \int_{\Gamma} \frac{1}{\mathcal{C}^2} \frac{\partial{\mathcal{C}^2}}{\partial {\bar{\eta}_1}} d \bar{\eta}_1
\label{theta}
\end{eqnarray}
and using the explicit form of the wavefunction it can be written as
\begin{eqnarray}
 \theta = - \textrm{Im} && \left[ \frac{1}{q} \sum_i \int_\Gamma \frac{d \eta_1}{\eta_1 - w_i}  + \frac{1}{q}\int_\Gamma \frac{q-1}{\eta_1 - \eta_2} d\eta_1 - \right. \nonumber \\ &&  \left. \frac{1}{q} \sum_i \int_\Gamma \frac{d\eta_1}{\eta_1 - z_i}  +
 \sum_i \int_\Gamma \frac{\left< n_i \right>}{\eta_1 - z_i} d \eta_1 \right] .
\label{theta1}
\end{eqnarray}
Observe that the first three members involve analytical functions and they can be computed easily.
Using Eq.~\eqref{theta1},
\begin{eqnarray}
\theta_a - \theta_b = i \left( 2\pi i \frac{q-1}{q}  + \sum_i \int_\Gamma \frac{\left< n_i \right>_a - \left< n_i \right>_b}{\eta_1 - z_i} d \eta_1 \right), \nonumber \\
\label{theta2}
\end{eqnarray}
where $\left<n_i\right>_a$ ($\left<n_i\right>_b$) is the density on the $i$-th lattice site with the quasihole at position $\eta_2^a$
($\eta_2^b$), as shown in Fig.~\ref{fig8}. Note that from the numerical results of Sec.V (see Fig.~\ref{fig5}) the function
$\left< n_i \right>_a - \left< n_i \right>_b$ in \eqref{theta2} is zero at every lattice site except around the positions $\eta_2^a$ and $\eta_2^b$, i.e.
\begin{eqnarray}
\left<n_i\right>_a - \left<n_i\right>_b = -f^a(z_i-\eta_2^a) + f^b(z_i-\eta_2^b),
\label{dens_approx}
\end{eqnarray}
with $f^a$ ($f^b$) some analytical function, nonvanishing only around $\eta_2^a$ ($\eta_2^b$), and such that $\sum_i f^{a(b)}(z_i-\eta_2^{a(b)})=(q-1)/q$. Finally, from Eq.~\eqref{dens_approx} and Eq.~\eqref{theta2} it is clear that $\theta_a - \theta_b = 0$ and therefore we obtain from Eq.~\eqref{alpha} that $\gamma=(arg(M_a) - arg(M_b))/2\pi=(q-1)/q$ and thus the quasihole anyonic statistics can be read directly from the monodromy and it remains invariant along the lattice-continuum limit interpolation.

\section{Magnetic field}\label{sec:magfield}
The Berry phase acquired when a quasihole moves around a closed loop while all other quasiholes are far away can be interpreted as an Aharonov-Bohm phase of a charged particle in an effective magnetic field $B$, i.e.\
\begin{eqnarray}
\theta =-i\ln(M)-(p_1 e/q) (2 \pi h) a \iint B(x_1,y_1) d x_1 d y_1 . \nonumber \\
\label{eff_B}
\end{eqnarray}
In the lattice limit the effective magnetic field felt by a quasihole in a unit cell is not completely uniform \cite{nielsen15}. However, in this section we show that it approaches the uniform value $B \rightarrow -h/(ea)$ (note that the magnetic flux through a unit cell is $-h/e$) as we approach the continuum limit. Using \eqref{eff_B} and the divergence theorem on \eqref{theta} the magnetic field, $B$, can be related to the occupation number $n_i$ \cite{nielsen15} (in natural units $h=e=1$ and taking $a=1$),
\begin{eqnarray}
B=-\frac{q p_{\eta_1}}{\pi} \sum_{i,j} \frac{ \left< n_i n_j \right> - \left< n_i \right> \left< n_j \right>  }  { (\eta_\alpha - z_i)
(\bar{\eta}_\alpha -\bar{z}_j) }; \quad \alpha=1,2
\end{eqnarray}
where again we split up a particle in two quasiholes at position $\eta_1$, $\eta_2$, far away from each other, and with charges $1/q$ ($p_{\eta_1}=1$) and $(q-1)/q$ ($p_{\eta_2}=q-1$) respectively.

\begin{figure}[htb]
\includegraphics[width= 4cm, height=3.5cm]{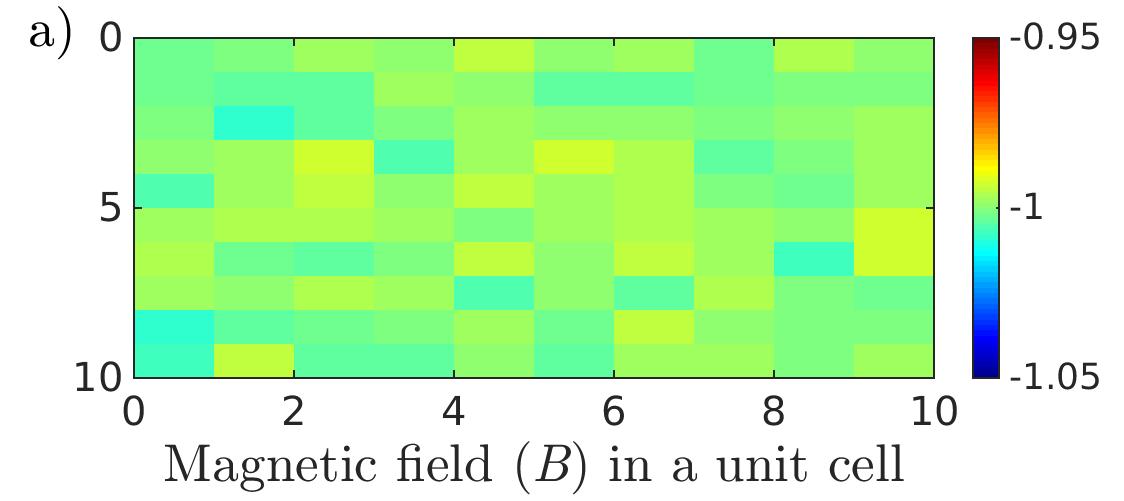}
\includegraphics[width= 4cm, height=3.5cm]{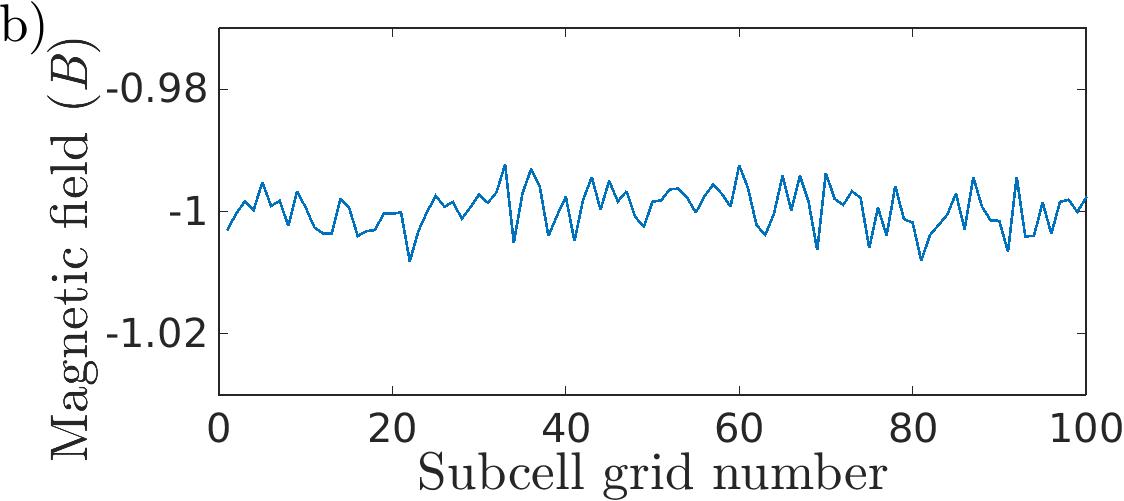}
\caption{{\small a) Effective magnetic field, $B$, felt by a $\nu=1/3$ quasihole with charge $1/3$ in a $10 \times 10$ grid (as explained in the main text) of a single unit cell, for the $3 \times 3$ subcell case. b) One-dimensional version of a) for the dependence of $B$ in every subcell grid. Note that the magnetic field is almost uniform with small oscillations around the uniform value -1 (in units $h=e=c=1$ and $a=1$).}}
\label{fig6}
\end{figure}

In Fig.~\ref{fig6} we show the dependence of the magnetic field felt by a $\nu=1/3$ quasihole with charge $1/3$ for the $3 \times 3$ case.
We divide a unit cell in a $10 \times 10$ grid, and in Fig.~\ref{fig6}a we plot the two-dimensional dependence of $B$ on the grid.
Fig.~\ref{fig6}b is a one-dimensional version of Fig.~\ref{fig6}a and shows the dependence of $B$ in every subcell of the grid
to make more clear that the magnetic field is almost uniform with small physical fluctuations (we have checked that the Monte Carlo errors are smaller than the fluctuations) of order $0.001$ around the mean value $-1$.
Finally, we have checked that similar results hold for the quasihole with charge $2/3$ and for both quasiholes in the $5 \times 5$ subcell case, though in this case the Monte Carlo errors are bigger and it is more difficult to distinguish the physical magnetic field fluctuations from the error bars.
\\
\section{Conclusion}\label{sec:conclusion}
In conclusion, we have proposed how one can use CFT to construct both lattice and continuum models with Laughlin-like ground states and to interpolate between the two limits. We have also shown that the topological properties of the models are as expected and computed the size of the quasiholes.

An interesting feature of the models is that both the Hamiltonian and the unique ground state are known analytically, both with and without quasiholes. This allows us to compute a number of relevant properties easily with Monte Carlo simulations. It also allows us to show analytically that the braiding properties of the quasiholes are as expected if the quasiholes are screened. The models hence remain within the same topological phase for all situations, where screening occurs.

The models are also interesting, because they allow us to compare the Hamiltonian obtained from CFT in the continuum limit with the usual delta function interaction Hamiltonian for the Laughlin states. The Hamiltonians turn out to be different, so that the CFT approach provides a different set of models displaying FQH properties.

Finally, we note that the approach presented in this paper can also be used to interpolate other FQH states constructed from CFT between the lattice and the continuum limit. This is so because the background charge always appears in the same way in the CFT correlators used to construct the states.

\begin{acknowledgments}
The authors would like to thank J. Ignacio Cirac for discussions. This work has been supported by the EU project SIQS.
\end{acknowledgments}

\bibliography{bibfile1}

\end{document}